A mechanism of synaptic clock underlying subjective time perception.

Bartosz Jura

Nalecz Institute of Biocybernetics and Biomedical Engineering, Polish Academy of Sciences, 02-109, Warsaw, Poland

Correspondence: bjura@ibib.waw.pl

**Abstract**

Temporal resolution of visual information processing is thought to be an important factor in predator-prey interactions, shaped in the course of evolution by animals' ecology. Here I show that light can be considered to have a dual role of a source of information, which guides motor actions, and an environmental feedback for those actions. I consequently show how temporal perception might depend on behavioral adaptations realized by the nervous system. I propose an underlying mechanism of synaptic clock, with every synapse having its characteristic time unit, determined by the persistence of memory traces of synaptic inputs, which is used by the synapse to tell time. The present theory offers a testable framework, which may account for numerous experimental findings, including the interspecies variation in temporal resolution and the properties of subjective time perception, specifically the variable speed of perceived time passage, depending on emotional and attentional states or tasks performed.

**Keywords**

**1. Introduction**

From everyday life we know that subjective perception of time is not constant and time might appear to flow slower or faster, depending, for example, on the emotional or attentional state or the task being performed (Dirnberger et al., 2012; Hagura et al., 2012; Lake et al., 2016). However, the exact physiological mechanisms that may underlie this phenomenon remain elusive.

It is generally acknowledged that time is not another sensory modality on its own, but a temporal dimension is inherent to perception of sensory stimuli, execution of motor tasks as well as various aspects of cognitive processing. Accordingly, the recent view is that there is no single, centralized clock in the brain, dedicated to performing any operations related to timing, regardless of the nature of a task. Instead, imaging data show that distributed brain areas are involved in time-keeping, which suggests that specialized areas have their own built-in timing mechanisms, that may vary among systems (Fortune and Rose, 2001; Buhusi and Meck, 2005; Díaz-Mataix et al., 2014; Muller and Nobre, 2014; Gouvêa et al., 2015). Consequently, timing has been proposed to be a direct result of the dynamical nature of neural activity and plasticity of the nervous system, which leads to the evolution of neural activity patterns over time. Time could thus be encoded in the progression of neural activity patterns through a phase space (Paton and Buonomano, 2018).

Importantly, timing of intervals in the sub-second range is thought to be automatic and not cognitively controlled, and to involve different brain structures, as opposed to timing over longer timescales (Lewis and Miall, 2003).

The studies of perception of temporal intervals over short timescales, particularly during exposure to stimuli coming from a single sensory modality, that is, for example, during observation of a dynamically changing visual stimulus, suggest that estimates of duration depend on the number of changes (i.e., bits of information), which are subjectively perceived in a unit time, rather than reflecting an objective duration of the interval (Kanai et al., 2006). Namely, when a higher number of distinct visual events is detected over a temporal window of a given length, the perception of this interval is more fine-grained and it seems to last longer. Consequently, the corresponding time flow seems to slow down, giving an impression of time dilation. Conversely, a lower number of events detected results in a more course-grained perception of the interval, which therefore seems to last shorter and the time flow appears to speed up. Accordingly, it has been proposed that estimates of duration correspond to the expenditure of neural energy used to process the stimulus (Eagleman and Pariyadath, 2009). Therefore, it appears that, in simple tasks limited to a single sensory modality, provided that duration of the interval which is to be timed is not compared to any reference duration values, such of stimuli coming from a different sensory modality or such stored in memory, the number of events detected can be studied as a proxy of subjective time perception.

Within the visual modality, a phenomenon which allows to directly measure the number of events detected in a unit time and, moreover, compare it in different animal species, is related to the property of the visual system termed Critical Frequency of Flicker Fusion (CFF). CFF determines the highest frequency of a flickering light stimulus, or the corresponding minimum interstimulus interval, which allows to distinguish subsequent light impulses as separate events. Frequencies of flicker above this value cause the impulses to be fused and perceived as a continuous stimulus. The exact value of fusion frequency depends on multiple factors, e.g., light wavelength or intensity or background lighting conditions, with the CFF usually specifying the highest threshold value found in any condition, representative for a particular species (Umeton et al., 2017). It is assessed using electroretinography (Frank, 2000) or, in animals, in behavioral paradigms in which animals are rewarded for responding to a stimulus, the flicker of which they can perceive (Lisney et al., 2011), and, in human, using psychophysical tests in which subjects report the critical frequency (Hardy, 1920). The phenomenon of CFF has been attributed to the temporal integration of signal in the retina and downstream pathways of information processing (Nardella et al., 2014). It is used as a measure of temporal resolution of visual information processing, representing a maximum number of bits of information that can be detected in a unit time. Higher CFF values correspond to higher resolutions, as more information is absorbed over a temporal window of the same length and more rapid changes in the stimulus can be detected, as opposed to lower CFF values, when information is integrated over longer time windows. Therefore, the value of CFF might be considered a measure of subjective time perception, as resulting from the processing of visual input.

The value of CFF was shown by Healy et al. (2013) to decrease with body size and increase with mass-specific metabolic rate of animals. Since body size affects the inertia and thus the maneuverability of animals, it was therefore proposed that CFF and, consequently, temporal perception, is an important factor in predator-prey interactions. Namely, given a sufficient level of their maneuverability and speed of movements, predators with higher CFF are able to precisely detect changes in the movement trajectory of a fast-moving prey and follow its steps and, conversely, prey with higher CFF might avoid being caught and escape predation by being able to see predator's movements in a "slow-motion" and react earlier (Figure 1). Therefore, the level of maneuverability determines whether it is worth to invest in a high-resolution visual system and hence it might have been a significant evolutionary constraint upon shaping the CFF in different species. It follows that the CFF can be studied as a functional trait which influences predator-prey interactions and which is shaped in an adaptive evolutionary-ecological game (Schmitz, 2017).

Here I show that CFF, as a temporal resolution of external sensory data processing, when considered from the perspective of predator-prey interactions, appears to be a result of behavioral adaptations realized by the nervous system. I derive it as a special case of a general limitation of temporal resolution of information processing, and I propose an underlying mechanism of synaptic clock, with every synapse having its characteristic time unit, determined by the persistence of memory traces of synaptic inputs, which constitutes a form of temporal filter, and which is used by the synapse to tell time. I also provide suggestions on how operations related to implicit interval timing, as emerging from the mechanism of synaptic clock, may be realized at the network level.

This viewpoint offers a testable framework in which numerous experimental findings can be considered, including the interspecies variation in the temporal resolution and the properties of subjective time perception in humans, specifically the variable speed of perceived time passage.

**2. Flicker fusion effect in predator-prey interactions and a dual role of light**

In the situation of predator-prey interaction an impulse of light can be considered 1) a source of information which guides animals' motor actions, and 2) an environmental feedback for those actions.

Consider an idealized situation of a predator chasing its prey, as depicted in Figure 2A. When the predator (square) initially detects the moving prey (circle) due to the visual information received with the first impulse of light (lightning bolt) (Figure 2A1), it starts to move itself (gray dashed arrows) and after every step it receives an update on the prey's position, in the form of another impulse of light, and so it can update its predictions and adjust its previous movement trajectory (Figure 2A2). Eventually, when the predator makes its final move and catches the prey, it gets the prey itself as a feedback for its motor action along with the last impulse of light (Figure 2A3). Thus, an analogy can be drawn here between the light and the prey, both serving as a feedback used to evaluate the predator's actions.

Therefore, the rate of processing of visual information (i.e., the CFF) should be adequate for the relative speed of movement of the predator and the prey, in order to allow the predator to catch its target (or, the other way round, the prey to escape predation). Specifically, a time interval between two impulses of light that can be distinguished as two separate sensory stimuli should be proportional to the interval between a visual information-guided action and the corresponding feedback for that action (in a given step of the "simulation", e.g., the one in Figure 2A2). If this interval is too short and the predator detects another sensory stimulus before its spatial position relative to that of the prey changes, then such information is redundant and the energy used to process it will be wasted. Conversely, if the interval is too long and the predator fails to detect another sensory stimulus after the relative position changes, it will not know what was the outcome of its action. It follows that the averaged interstimulus interval of CFF should be fine-tuned in order to give organisms a survival advantage. Now, we will consider the constraints that limit the value of this interval.

The shortest possible interval between a motor action and a feedback is determined by the highest possible speed of predator-prey interaction and information transfer, which is the speed of light. Therefore, the highest possible value of CFF, giving the organism a virtually perfect (i.e., infinitely high) temporal resolution of information processing, should be developed when light itself is a "prey" and a feedback for motor actions, for example when it is utilized by an organism to produce nutrients in photosynthesis, which is actually the case in the phenomenon of phototaxis.

2.1. The dual role of light in phototaxis

Phototaxis is a phenomenon found both in prokaryotes and eukaryotes, in which organisms can move along a light gradient. It has been found, among others, in photosynthesizing organisms (for a review, see Jékely, 2009).

Recently, it was shown by Schuergers et al. (2016) that one of such organisms (unicellular cyanobacterium *Synechocystis* sp. PCC 6803) acts as a micro-lens and can precisely sense the direction of light and move directly towards its higher intensity, in order to obtain, as an effect, more light energy for the photosynthesis. Thus, in this organism light serves both as a sensory cue which guides its motor actions, and a "prey", being an environmental feedback for those actions (Figure 2B). As soon as the organism makes a move towards the light (Figure 2B1), a brighter ray of light can reach its thylakoid photoreceptors and nutrients be produced in photosynthesis (Figure 2B2). Thus, the time interval between the execution of a motor action and the delivery of feedback is proportional to the speed of light in the medium.

Therefore, in order to be useful and allow the organism to react immediately to any changes in the illumination direction, the temporal resolution of sensory information processing, determining the minimal interval between two subsequent visual stimuli which are not to be fused into one, should also be proportional to the speed of light in the medium. However, since the molecular signals of the cascade triggered by the sensory input and leading to the execution of a motor action cannot decay faster than with the speed of light, the actual resolution will be lower and limited by the decay rate of that molecular trace left by the input.

In this case of phototaxis, the goal of organism's actions is itself used as a sensory cue which guides the actions. Thus, the relation between the sensory information and the goal is constant and hence the information-guided actions do not need to be evaluated and adjusted according to their outcomes. Therefore, the molecular signal might decay spontaneously, immediately after it triggers a motor action and there is no need to sustain it until a feedback is delivered. This appears to hold in any other cases where movements are simple reflexes and where the relation between sensory information and goals does not change, also when the relative speed of movement of an organism and its goals is lower than in the case of phototaxis. For example, in chemotactic organisms, which can detect and move along the concentration gradient of various chemical substances (for a review, see Wadhams and Armitage, 2004). There, similarly, a target substance is used as an action-guiding sensory cue and thus the relation between them is constant (or dependent only on a temporary internal state of the organism, e.g., its current metabolic demands). Therefore, the input signals also do not need to be sustained until a feedback delivery (or except only to detect instantaneous changes in the concentration gradient). Hence, there are no constraints on the minimal time of persistence of input signals and therefore the value of temporal resolution does not have a strong upper bound.

However, the situation is different in organisms which can use neutral information with changing relation to goals to guide their motor actions, which is the case in organisms with a nervous system.

**3. Synaptic plasticity and behavioral adaptations**

The evolution of the nervous system has separated the channels of motor output from the channels of sensory input with additional, intermediate stages of information processing. It enabled organisms to use various types of information with complex relation to goals to guide their motor actions. Importantly, neutral information can be used by them, and the relation of information to goals can change. Learning of abstract, statistical relations between neutral information and goals, based on the outcomes of the actions triggered by such information, is realized through the activity-dependent plasticity of the nervous system, which leads to behavioral adaptations. The most well-studied aspect of such plasticity is the activity-dependent synaptic plasticity (Sweatt, 2016). The view of behavioral adaptation as the primary function of synaptic plasticity is supported by the features of the nervous system. Namely, any act of synaptic transmission eventually affects the execution of motor actions, due to the polarity of synaptic connections and the directionality of transmission of effective signals in the nervous system. Ultimately, sensory inputs are received through some dedicated channels, and motor outputs are generated through others. The existence of recurrent connections does not contradict this directionality, as any circuit of recurrently connected cells must eventually receive an extrinsic input and project an output. Any activity-dependent change in synaptic weights should therefore be beneficial and not harmful for the organism as a whole, that is, it should result in actions the performance of which is adjusted to given situations.

This utilitarian nature of the activity-dependent synaptic plasticity is reflected in the three-factor rule of synaptic plasticity, which suggests that not only co-activation of pre- and postsynaptic cells is required for an enduring synaptic modification to occur, but also the influence of neuromodulation, with dopamine (DA) playing a key role in this respect in the mammalian brain (Lisman et al., 2001). DA has been shown to be required for the protein synthesis-dependent late phase of Long-Term Potentiation (L-LTP) in distributed brain areas (Otani et al., 2015; Broussard et al., 2016). DA might exert its permissive effect on plasticity even when it is delivered with a delay (Lisman and Grace, 2005) and, moreover, it can bidirectionally affect the synaptic change (i.e., allowing for L-LTP or late-phase Long-Term Depression (L-LTD) (Huang et al., 2004), or, depending on the concentration, transforming a potential L-LTP into L-LTD and *vice versa* (Reynolds and Wickens, 2002; Thivierge et al., 2007; Ruan et al., 2014)), and thus it can be viewed as an internal substitute of environmental feedbacks for synaptic activity.

Importantly, since organisms live in spatially extended environments, environmental feedbacks are never delivered immediately after a synaptic activity, with the speed of feedback delivery limited by the speed of light. Sometimes, a relevant feedback may be very distant in time from an information-guided action. As a result, organisms need to adjust their behavior based on delayed feedbacks, and different types of information-guided actions tend to have different temporal distance to the action-evaluating goals.

**4. A mechanism of synaptic clock**

4.1 The persistence of memory traces of synaptic inputs

Consider an idealized organism (O1) with a simple neural circuit, as depicted in Figure 3A1, and the synaptic connections between a presynaptic input neuron and two postsynaptic output neurons. Assume that the input neuron is a higher-level unit which detects a specific type of places in the environment, based on a combination of visual cues (crossroad), and the output units innervate limbs which can change direction of the organism's movement, turning the body either left or right. Assume further that the organism is moving forward at a constant speed in a medium and has limited energy resources, sufficient only to reach the nearest energy supply (polygon). When the organism arrives at the crossroad then, in order to survive, it needs to enter the arm with the food and to do so it has to perform a motor turn in the appropriate direction. Initially, the weights of the synaptic connections are random. Then, if the sensory unit does not activate the correct motor unit and the organism enters the empty arm, it will not reach the food and die. If the organism does perform the correct motor turn and receives an environmental feedback for that action in the form of the food reward (Figure 3A2), and it leads to the strengthening of the synapse the activity of which has led to this beneficial outcome, the survival chances of the organism will increase. However, if the organism performs the correct action but the activated synapse is not strengthened after receiving the feedback, the chances of performing the correct action when being next time in similar situation, and thus of survival, will stay on the same level as with the random set of synaptic weights.

Since the feedback delivery is not immediate and, on the other hand, synaptic activity can be transient and short-lasting, the synaptic input needs to leave a mark on the synapse so that it can be reinforced when a relevant feedback is delivered. From now on, we will refer to this type of mark as the synaptic *tag*.

4.2. Duration of a synaptic tag

If such an activity-dependent tag does not persist for long enough and decays before the organism can reach the food reward, then, even after performing the correct action and receiving the feedback, the synapse will not be modified and the final result will be the same as there was no synaptic tag at all. Conversely, if the duration of a tag is too long, the activated synaptic connection can be altered due to further environmental feedbacks, unrelated to the action triggered by the particular input stimulus and the activity of the synapse. Eventually, the modification of the synaptic weight can get reversed and, as a net effect, the activity of the synapse in the given circumstances will leave no trace in the network and, yet again, the situation will be analogous to the lack of plasticity. However, when the duration of the tag is properly adjusted and it persists just until the environmental feedback for the given synaptic activity is delivered, the connection can be reinforced effectively. Subsequently, when the organism finds itself again in a similar situation, it can make the right decision, based on its previous experience, on how it behaved and what outcome that action had brought.

Eventually, the only property that distinguishes an organism that will survive from any other individuals is the duration of the synaptic tag being fine-tuned for the type of information processed in the given synapse, which allows for an effective act of synaptic plasticity to occur (Figure 3A3).

4.3. Different types of synapse

Consider now another organism (O2), as depicted in Figure 3B1, analogous to the O1, and the synapses between an input and two output neurons, the functions of which, however, differ from those of the O1. The sensory unit detects food portions that the organism is approaching. Short-lasting activation of one of the motor units leads to opening of the mouth and thereby enables the organism to eat the food when it reaches its position (Figure 3B2). On the other hand, the activation of the other motor unit results in execution of another, task-irrelevant action, e.g., moving a limb, and consequently, when the organism reaches the position of the food, the food gets bounced off its body and, as a result, it will die from a lack of energy. Therefore, for the type of synapse in this organism, the time interval between a synaptic activation and the delivery of relevant environmental feedback is shorter than in the case of the O1. It follows that the synaptic tag, in order to be advantageous, should also have shorter duration (Figure 3B3).

4.4. Synapse type-specific tag duration

Therefore, we argue that every synapse, belonging to a class of synapses processing a given type of information, develops a mechanism of synaptic tag of a specific duration, proportional to the expected value of the time interval which separates synaptic activations from the delivery of relevant environmental feedbacks.

4.5. Synaptic CFF

The synaptic tag can be viewed as a short-lasting memory of the synaptic event that led to its creation and, consequently, the rate of tag decay can be viewed as a rate of forgetting about that event. Similarly, the fusion of sequential visual stimuli can be viewed as a result of short-lasting memory of single stimuli (also known as visual persistence (Hardy, 1920)) and the CFF can be viewed as a rate of forgetting about those stimuli. Therefore, by way of analogy, a synapse can be considered as a "sensory" unit, which receives external signals of a certain type and sustains the memory of the input for a characteristic period of time, determining its temporal resolution of information processing. Within this time interval, the incoming inputs will be fused, and the same feedbacks will be applied to them, and thus the interval will constitute a synaptic "CFF".

4.6. Synaptic clock

Therefore, since a default duration of such a synaptic tag is constant, it constitutes a time unit which can be used by the synapse to tell time. Hence, synapses might work as clocks and, since tag duration is synapse-specific, every synaptic clock will be telling time in its own unit.

4.7. Synaptic clocks and anticipation

The setting of a synaptic tag, by an event itself neutral, can be viewed as an act of prediction that a feedback for this synaptic activity will be delivered in the future, within the time interval specified by the tag duration. Thus, the operations of a synaptic clock, namely the decay of the synaptic tag magnitude, can be viewed as a passive process of goal anticipation.

When an organism comprises multiple synapses, each of them processing a different type of information (e.g., a composition of the O1 and the O2), then every synapse will have a tag mechanism with a different individual time constant. However, the goals pursued or avoided by the organism, having a significance for its survival, influence the organism as a whole and thus are common for all of its components. None of the organism's synapses "knows" what goal is being pursued at the moment. The only thing any of the tagged synapses "knows" is how far the goal is (or how fast it is being approached), as indicated by its individual time unit. Therefore, regardless of what the current goal is, the activity of every synapse indicates its own, expected time remaining to the goal delivery. Thus, all synaptic clocks are moving with respect to a common frame of reference, a current goal, which is also the frame of reference for the organism as a whole. Consequently, the temporal resolution of the organism will be proportional to the weighted sum of activity of all the synaptic clocks that it contains.

Eventually, the emerging sense of the speed of time flow is inherently implicit and does not rely on intentional, cognitively driven referring to the contents of memory and measuring a time interval that has passed since some event in the past. It stems rather from a continuous process of goal anticipation and thus is related to an event expected to occur in the future.

**5. Attention**

Attention can be defined to be manifested through an increased activity rate of presynaptic cells in some neural population. Thus, attention leads to the elevated stimulation of the postsynaptic cells and, eventually, setting of synaptic tags with high, initial magnitudes. As a result, a short-lasting memory of inputs processed by the activated synapses is stronger, relatively to that in other neural populations, and has larger contribution to determining what the currently pursued goal is. Therefore, the temporal perception will be more dependent on the pool of synapses in such a highly active region. Ultimately, the overall temporal perception of an organism might be reduced to the dynamics of synaptic clocks with different time units.

**6. Planning**

As stated above, a synaptic tag indicates an expectation that a feedback for a given synaptic activity will be delivered in the future. Feedbacks for actions of different synapses will be expected to be more or less distant in time, proportionally to the durations of their synaptic tags. Thus, different goals, related to different types of information processed by different synapses, will be expected at different time points in the future. Since goals which are less distant tend to be more likely to be successfully delivered, they should have a stronger impact on a decision-making process. It follows that a "real" planning is limited by the duration of the longest synaptic time unit. In contrast, planning with regard to goals in a more distant future appears to be a mere abstraction, derived from the activity of an existing pool of synapses with shorter time units.

## 7. A global distribution of synaptic clocks

7.1 Synaptic tag as a dynamical process and the issue of timescale

The mechanism of synaptic tag, as proposed originally within the framework of synaptic tagging and capture, demonstrates that a weak synaptic stimulation can leave a protein synthesis-independent tag on the synapse, so that proteins synthesized non-locally in the cell body, possibly also due to other unrelated events, can find the synapses that have been activated recently and secure their long-lasting modification (L-LTP) (Frey and Morris, 1997).

The molecular nature of a tag might depend on the synapse type (as it depends on the process involved (i.e., LTP or LTD) and localization within a dendritic tree (e.g., basal or apical dendritic compartments in hippocampal CA1 neurons)) (Martin, 2002; Sajikumar et al., 2007; Okamoto et al., 2009). Consequently, the central mechanism of the synaptic clock may be realized in different synapses by different molecular underpinnings. Moreover, different synaptic clocks within a cell will be presumably interrelated, due to intracellular signaling and diffusion of tagging molecules, especially within synaptic compartments (Sajikumar et al., 2007).

The synaptic tag is identified as an early phase of LTP (E-LTP), that is, a synaptic change which does not require protein synthesis, transformed into the late phase through the action of captured proteins. It has been proposed to underlie a short-lasting memory of behavioral events (Ballarini et al., 2009; Wang et al., 2010). The same mechanisms and molecules have been implicated both in the synaptic tagging and the E-LTP, notably, among others, the $Ca^{2+}$/calmodulin-dependent protein kinase II (CaMKII) (Moncada et al., 2011; Lisman et al., 2012). In hippocampal neurons, a synaptic tag has been shown to persist for about 1 hour, as proteins synthesized this long after an initial synaptic event can effectively reinforce the change of the synapses involved (Lisman et al., 2001; Ballarini et al., 2009). However, a molecular cascade triggered by a synaptic stimulation appears to consist of multiple stages, all of which are required for a synaptic modification to occur and which, importantly, have different durations. For example, CaMKII autonomous activity (which is limited to single dendritic spines (Lee et al., 2009)) lasts only for about 1 min., whereas the state of its autophosphorylation persists for longer time periods (Lengyel et al., 2004).

Therefore, based on these observations, a synaptic tag can be viewed as a dynamical process, comprising a sequence of molecular processing stages, covering multiple timescales. Thus, the time unit of a synaptic clock should be proportional to a total duration of the corresponding molecular cascade but it might as well consist of multiple stages and be determined primarily by the effect this cascade has on the cell and circuit activity and, consequently, behavior.

7.2. A distribution and action of synaptic clocks

An analogy for such sequential processing of information can be found in the visual system, where a short-lasting visual persistence of stimuli, being an early phase of information processing, is quickly transformed into a medium-term memory, which helps constructing the entire visual scene from a sequence of visual elements, viewed one by one (Melcher, 2001). Therefore, the other way round, the persistence of visual stimuli, determining the CFF, could be considered as being due to an initial stage of synaptic tags with very short time constants.

Consequently, the global distribution of time units of synaptic clocks in the brain could be proposed, in relation to sensory and motor extrema (Figure 4). The values of time units will correlate with the level of abstraction of information processed in a given synapse, which, in turn, is related to the relative, average distance of the synapse from the sensory and motor regions. Then, when a purely sensory information is processed or an automatic, reflex motor action is executed, without involvement of higher-order, associative brain areas, synaptic clocks with shorter time units will be activated, which should result in more information processed in a unit time and the impres-sion of a more fine-grained experience, as if the time flow was slowing down. On the other hand, when more abstract, for example spatial, information is processed, the brain areas containing synaptic clocks with longer time units will be active, e.g., the hippocampus, which should result in fewer bits of information being processed in a unit time and the impression of time flow speeding up.

This distribution allows to make predictions, firstly, about the direction and magnitude of time perception distortions depending on the task performed and attention directed to different aspects of the task (e.g., sensory modalities involved), and, secondly, about the interspecies differences in temporal perception, which should be the result of nervous systems having distributions of synaptic clocks with different time units.

**8. Sleep and systems memory consolidation**

The duration of a synaptic tag determines the maximum time interval between a synaptic activation and the delivery of an environmental feedback, which can effectively reinforce the synaptic change. Therefore, the lifespan of a simple organism (e.g., the O1, Figure 3A) is limited by the duration of the longest synaptic tag mechanism that it contains. Specifically, if feedback for a given synaptic action does not arrive before the tag decays, the organism will never know what was the outcome of that action and, as a result, it will not be able to adjust its behavior, in order to adapt to the given type of situation. Moreover, with an increase in synaptic tag durations, synaptic inputs become integrated over longer time windows, so that in natural, non-uniform environments every input will be encoded with an ever decreasing weight. Eventually, it will impose a limitation on the maximum effective duration of a synaptic tag and thus on the lifespan of simple organisms.

However, environments and life events tend to be repetitive in the time domain, namely, for example, the number of different types of events that might happen to an organism between a sunrise and a sunset is limited. Therefore, an organism that would be able to generalize the knowledge could "repeat" its entire lifetime multiple times, using only the information acquired over a limited timespan. Thus, when an organism is able to slow down its vital processes and fall into a state of reduced activity, in order to limit its energy demands and possible dangers that it may encounter, and use this period to transfer the information, in an "unsupervised" way, to a system that is less plastic and can keep the memory traces for longer periods of time and generalize upon them, it gives an evolutionary advantage to such organism. Therefore, during the systems memory consolidation in sleep information will be transferred to be stored within a set of synapses with tag mechanisms so long, and thus so weakly plastic, that they do not encode any information which is available directly in the environment. Eventually, when information can be generalized upon, the distribution of tag durations might be shortened accordingly. It follows that the period of sleep-wake cycle should be proportional to the distribution of synaptic tag durations, and hence it should correlate with the temporal perception in different species. Therefore, this is a potential way to verify the present hypothesis, which, moreover, offers a link between two timescales (i.e., sub-second and circadian) thus far not thought to derive from a common underlying mechanism (Paton and Buonomano, 2018).

Sleep is often seen as serving the needs related to information processing in the brain, that is, consolidation of memory and synaptic downscaling (Buzsáki, 1996; Rothschild et al., 2017). On the other hand, sleep (and analogous states of inactivity) has been also postulated to be itself a form of behavioral adaptation, shaped by a combination of ecological factors and increasing the survival chances of animals (Siegel, 2009). The present viewpoint offers an integration of these two views on the sleep origin and function, as the need for memory consolidation will be determined by the distribution of synaptic tags durations which, in turn, is shaped by the animals' ecology and rate of interactions with their surroundings. This perspective is supported by observations that animals which do not sleep, or express a reduced amount of sleep, usually live in spatially unstructured and monotonous, event-poor environments (e.g., marine animals), lack natural predators (e.g., sharks, cavefishes) or use strategies which decrease their demands for processing new information, limit exposure to threats and thus give time needed for memory consolidation and maintenance of neural circuitry (e.g., schooling, turtles' shell) (Kavanau, 1998; Borowsky, 2018).

Moreover, the phosphorylation of CaMKII in neurons of the dorsal raphe nucleus has been implicated in the regulation of sleep-wake cycle in mammals (Cui et al., 2016), which may further support the hypothesis, given the role of CaMKII autophosphorylation in synaptic tagging and plasticity.

### 9. Brain imaging methods allowing for on-line monitoring of temporal perception dynamics

Techniques of functional brain imaging which measure signals of the oxygenated blood flow (with magnetic resonance, fMRI, or near-infrared spectroscopy), can indicate an increase in activity of particular brain regions. It has been postulated that they are sensitive primarily to the presynaptic activity, i.e., the activity of afferents to a given site and magnitude of synaptic release (Logothetis et al., 2001), which satisfies the definition of attention as used above. Therefore, these techniques might give a measure of the relative contribution of a particular neural population with a specific set of synapses to an overall, ongoing temporal perception, based on the presumed abundance of recently potentiated synapses with high-magnitude synaptic tags in regions which are activated. Indeed, there is evidence that increased activations of specific brain regions (i.e., right putamen, amygdala and insula), as indicated by fMRI, correlate with distortions of time perception (Dirnberger et al., 2012).

Techniques which measure field potentials generated by populations of neurons (electroencephalography (EEG) or local field potential recordings), as reflecting mostly synaptic currents (Haider et al., 2016), in addition to the interregional (amplitude of a presynaptic bursting activity, indicating the relative contribution of a region), might also indicate the intraregional, local synaptic dynamics (rate of a presynaptic bursting activity). Namely, when a given neural population is working in synchrony, these techniques can indicate the frequency of its oscillating activity and thus the rate of synaptic input to the postsynaptic cells, thereby showing the amount of information received by the synapses in a unit time. Importantly, when the interstimulus interval (i.e., oscillation period) is shorter than the time units of relevant postsynaptic clocks, the inputs will be fused, however, when the interval is longer, its value will determine the number of changes detected by a synapse in a unit time. For example: 1) a direct correlate of the CFF measure is found in EEG recorded over the visual cortex during observation of a flickering light stimulus, as Steady-State Visual Evoked Potentials, that is, field oscillations with a frequency corresponding to the frequency of the flicker (Kuś et al., 2013), which should correlate with alterations in the perceived speed of time flow, when different amount of visual information is detected in a unit time, depending on the frequency; 2) frequency of the theta rhythm in hippocampus correlates with the running speed of animal (Sławińska and Kasicki, 1998) and the running speed, in turn, should also correlate with the perceived speed of time flow, as the same amount of time is used to cover distances of different length; 3) the high-amplitude low-frequency oscillations in delta band (~1 Hz) recorded in the cortex during quiescent states (Buzsáki, 1996) appear to speed up the perceived time flow, as less amount of information is processed in a unit time and time intervals appear to be shortened.

### 10. Neuromodulation

Neuromodulation, especially by DA, can be viewed as a substitute of environmental feedbacks for synaptic activity (Schultz, 1997; Wang et al., 2010), required for the L-LTP (i.e., enduring synaptic change), as substantiated by the three-factor learning rule, but it also has an ongoing influence on the neural activity and might alter time perception (Paton and Buonomano, 2018), especially in emotional states. Specifically, it could exert its effect either by directly affecting the dynamics of tagging molecules, changing the magnitude and decay rate of synaptic tags, or indirectly, by modulating the neuronal excitability, presynaptically relative to a given site, and thus altering the rate of synaptic input. Such a DA action may explain the reported dependence of temporal perception on the midbrain DAergic neurons' activity (Soares et al., 2016) and its distortions in pathological states affecting the DA system, e.g., in Parkinson's disease patients (Mioni et al., 2018).

## 11. Network mechanisms of interval timing

The mechanism of synaptic clock suggests emerging network mechanisms, related to subjective time perception and implicit interval timing, to be distributed over specialized brain areas and not dedicated to the processing of time *per se* but rather timing being indistinguishable from the sequence of neural events during execution of an actual task.

It has been proposed that memory is allocated to specific cells and synapses through a set of interrelated processes, which include synaptic tagging, dendritic spine clustering and alterations of neuronal excitability, and that a neural event at one time point affects the probability of recruitment of the same specific neural substrate by subsequent events occurring in temporal proximity. Eventually, network activity drifts over time and representations of events occurring close in time are more similar to each other than to those of other, more distant events, a process which is thought to contribute to the formation of relational memory (Miyashita, 1988; Rogerson et al., 2014; Cai et al., 2016; Eichenbaum, 2017).

Importantly, setting of a synaptic tag, and thus E-LTP of the synapse, can alter excitability of the cell (e.g., by changing permeability of ion channels in synaptic compartments) (Campanac and Debanne, 2007) and it allows to see a decaying synaptic tag as a dynamical process affecting the ongoing circuit activity. Consequently, the evolution of a neural activity pattern might be, in part, due to changes in neuronal excitability resulting from the synaptic processes. Therefore, the mechanisms of interval timing which derive from such evolving patterns of activity may apply. For example, one could conceive of synaptic tags set on an inhibitory, signal-integrating neuron by an initial stimulation, at the onset of an interval which is to be timed. The subsequent decay of the tags will result in decaying excitability of this cell, and its decaying firing rate, elicited by a weak tonic input, will eventually result in the release of inhibition from its downstream targets. This will lead to a ramping activity of these downstream target neurons, like, for example, in the parietal cortex of a monkey, in which primary neurons increase their firing rates with time elapsed before the action onset, initiated when the firing reaches a certain level (Jazayeri and Shadlen, 2015). Such a mechanism could help explain the differential contribution of distinct brain regions, with different putative default durations of synaptic tags, to the timing of intervals depending on their length (Lewis and Miall, 2003).

## 12. Discussion

In the present study I have shown how the properties of subjective perception of time may arise in different animal species, and how they depend on the persistence of traces of previous inputs, with the synapse, and the proposed mechanism of synaptic clock, being a plausible site where such operations are performed.

Extrapolation of the CFF property, specifying a temporal resolution of information processing, to brain functions other than visual perception may be justified by the studies of visual imagery. Namely, they show in human, using fMRI, that an overlapping neural substrate is activated both by an actual sensory experience and its subsequent explicit imagery (i.e., recall) (Albright, 2012). Consequently, a detailed recall of past events is thought to be a reconstruction of previous states of the brain. Moreover, an implicit imagery is an integral part of any sensory experience, with sensory data being completed and interpreted in light of the contents of memory, accumulated over the lifetime, in order to form a meaningful mental scene. These observations suggest that direct sensory perception and, on the other hand, internal, autonomous mental states might be viewed as a continuum and a property analogous to the CFF, as strictly dependent on memory, will apply to any of them.

It is generally assumed that a higher temporal resolution of information processing should be associated with higher energy expenditure per unit time and should correlate with higher rates of neural activity and metabolism (more "effort" put into processing) and, on the other hand, with more neural resources used to represent the information. I argue, however, that the resolution is determined primarily by a process which is essentially passive, and that a higher resolution is a result of shorter persistence (i.e., faster decay) of memory traces, as substantiated by the putative decay of postsynaptic tags. Shorter time units allow for more flexibility of the neural activity and higher number of distinct events to be processed in a unit time (which leads to shorter reaction times (Hagura et al., 2012)), but a higher resolution is hypothesized here to be related to the processing of information which is simpler (i.e., less abstract, direct sensory data), and such information should require less neural resources to be represented. Therefore, an increase in the amount of information per unit time will be compensated by a decrease in the complexity of this information. Indeed, the temporal resolution is shown to increase not with net, but with a mass-specific metabolic rate (Healy et al., 2013).

The limited temporal resolution with which organisms can process information, as substantiated by the CFF interstimulus interval, implies a basic unit of time which separates any two events that can be effectively distinguished as different events, occurring at two different time moments, and which cannot be divided further by the organism into any shorter fragments. Therefore, one might say that for us a single time moment has some non-zero effective duration and cannot be reduced to a discrete point. This conclusion echoes what is postulated by some authors in regard to the nature of physical time. Namely, it is proposed that a moment of physical time might, in fact, have certain duration or "thickness" (see, for example, Smolin, 2015). Therefore, the viewpoint presented here may be informative, as a source of analogy, for the considerations of physical time. Specifically, one could conceive of going down, in a sense, from the organic to the non-organic matter, which obviously has no practical needs and therefore does not need to remember anything or develop any memory in the biological sense, but still it *interacts* with its surroundings through the electromagnetic and other physical forces. Thus, a parallel could be drawn, from the perspective of information transfer or rate of local changes of physical structure, to see whether for non-organic systems the time intervals between information-triggered "actions" and "feedbacks", and thus between consecutive time moments, go to zero or, rather, also have some non-zero durations.

**Acknowledgments**

I thank Tiaza Bem and Pierre Meyrand for their helpful comments on the manuscript.

**Declarations of interest**

None.

**Figures**

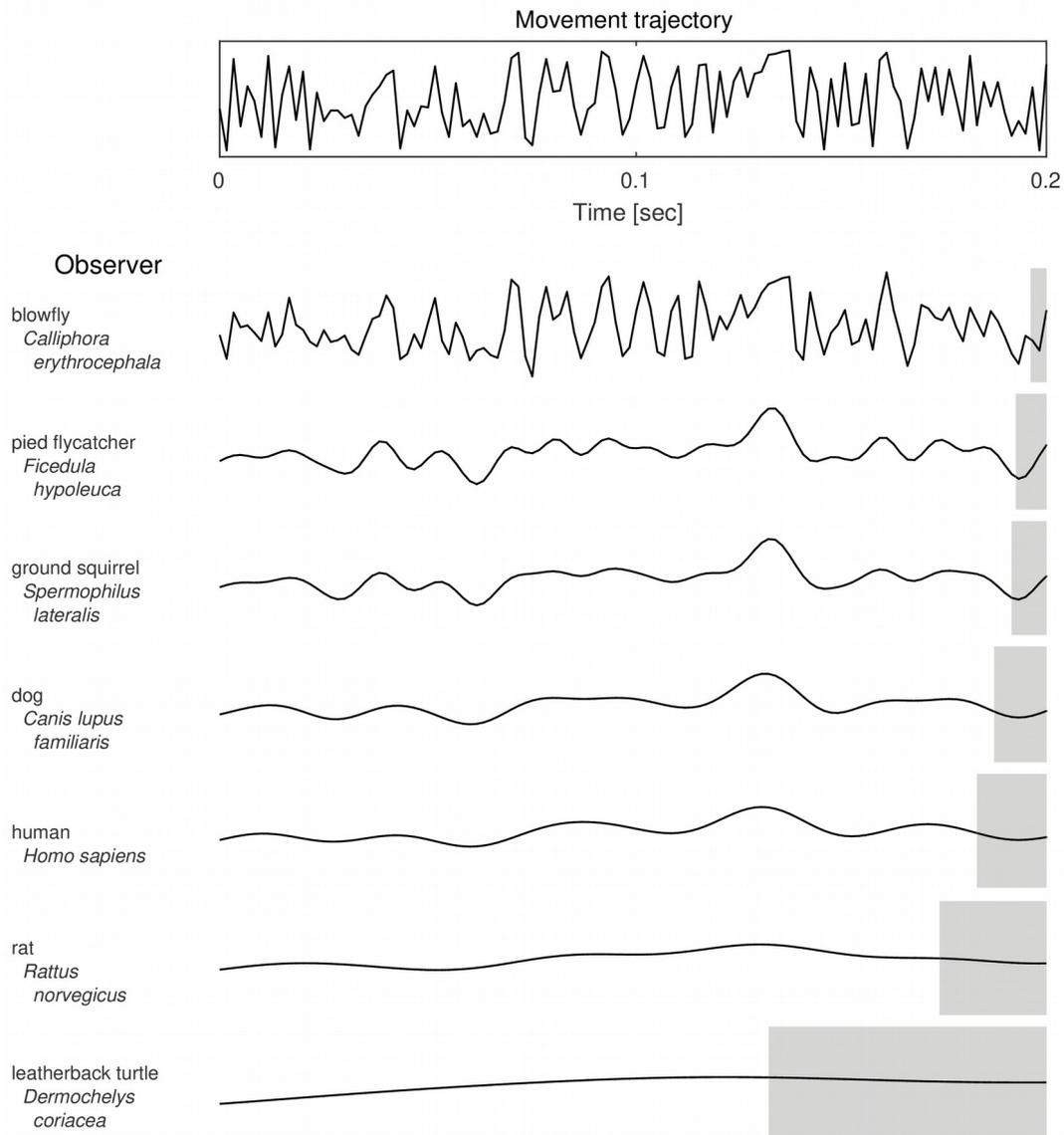

**Figure 1. The putative effect of the CFF value on the movement trajectory of another animal, as perceived by different species.**

With a decrease in the number of bits of visual information detected in a unit time (from the top to the bottom of the table), a perceived trajectory of the actual movement (upper panel) is more course-grained and hence the estimated position of the observed animal is less accurate. Grey shaded areas indicate relative lengths of the corresponding temporal windows of information integration, that is, intervals of duration equal to 1/CFF. Curves were drawn based on the CFF data taken from and referenced by Healy et al. (2013), except for the blowfly (Autrum, 1949) and the pied flycatcher (Boström et al., 2016).

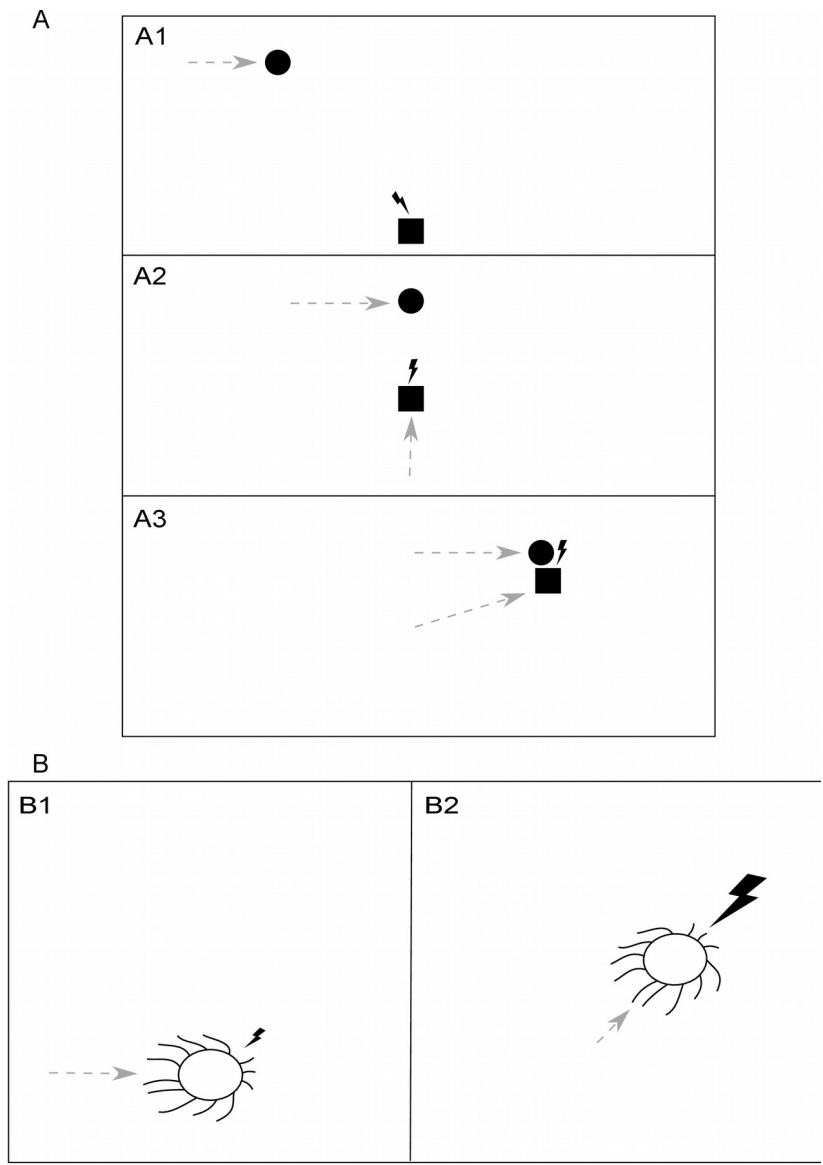

**Figure 2. The dual role of light as a source of information, which guides motor actions, and an environmental feedback for those actions.**

**(A)** In the predator-prey interaction (i.e., prey pursuit) an impulse of light (lightning rod) serves as (1) a sensory cue which indicates the position and movement of the prey (circle) to the predator (square) and guides its subsequent motor action, (2) a feedback for the predator's action (dashed gray arrow) and another cue, allowing the predator to update its information about the prey, and (3) a feedback for the predator's action, delivered along with (or being a substitute of) the prey itself. The temporal resolution of visual information processing, in order to be advantageous, should be proportional to time intervals between information-guided motor actions and the corresponding feedbacks delivery. Thus, it should be adequate to the speed with which the relative spatial position of the animals changes, that is, the relative speed of movement of the predator and the prey. **(B)** A direct manifestation of light being an environmental feedback is found in the phototaxis, where it is utilized, for example by cyanobacteria, in photosynthesis. An impulse of light, falling from a certain direction, is a cue which guides the motor action of the organism (1). The subsequent light impulse of higher intensity is a reward for the action, delivered after the organism makes a move towards the light (2). This example appears to constitute a limit for the temporal resolution of information processing, as it should be proportional to the speed of light in the medium and hence will be determined only by the rate of underlying molecular processing.

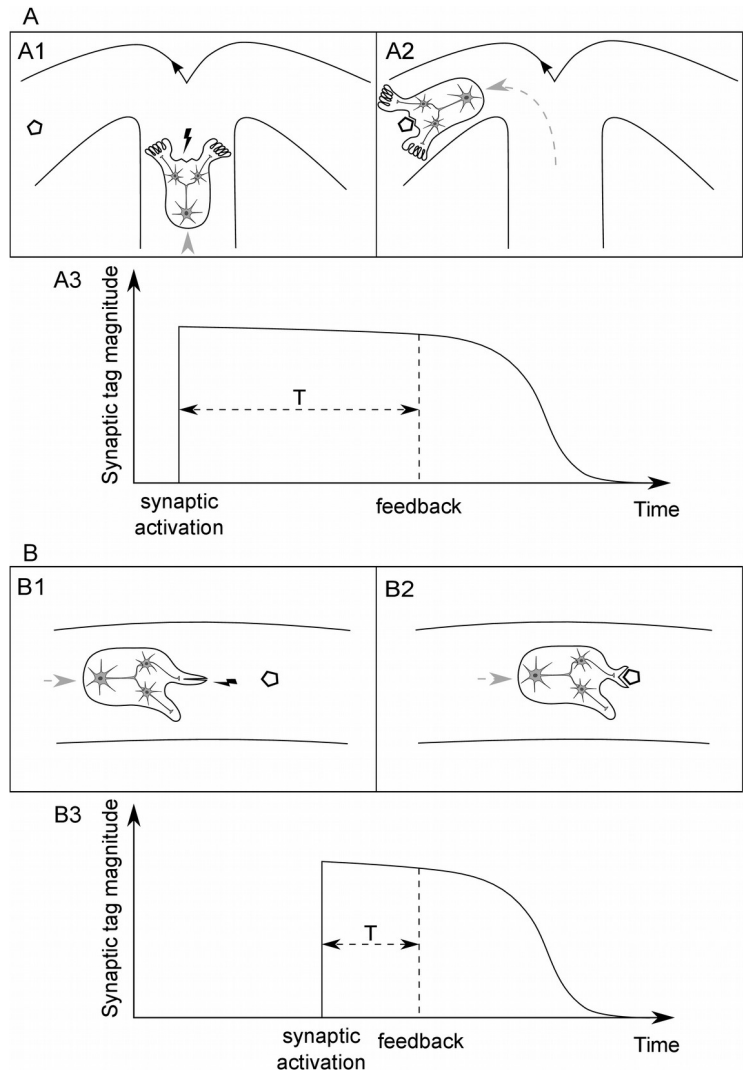

**Figure 3. The persistence of memory traces of synaptic inputs, and its dependence on the type of information processed by a given synapse and its temporal distance to goals.**

**(A)** (1) Cartoon depicts an idealized organism with a simple neural circuit, moving at a constant speed in a medium and having limited energy resources, the input neuron of which detects specific combinations of visual cues in the environment (crossroad) and activates one of the postsynaptic output neurons, which change direction of the organism's movement. In order to be modified appropriately, based on the outcome of its activity, an activated synapse needs to stay tagged until the position of the food (polygon), or the end of the empty arm, is reached (2). (3) A time course of the synaptic tag magnitude. The required duration of persistence of the synaptic tag will determine a characteristic time constant for this type of synapse. **(B)** (1) An organism analogous to that in (A) but the input neuron of which detects food portions being in close proximity and activates one of the output neurons, which leads either to opening of the mouth or execution of some other, inadequate, motor action. Here, the feedback for a motor action is delivered within a shorter time window than in (A) and thus the synapse needs to stay tagged for a shorter period of time (2). (3) This synaptic tag should decay faster than that in (A), and thus it will determine a shorter time constant for this type of synapse. Different types of synapse will have different, characteristic time units, shaped as the averaged temporal distance of a particular type of information-guided actions to goals. The actual values of time units will be shaped by organisms' speed of movements and rate of interactions with their surroundings. The synapse depicted in (A) might be an analogy to a synapse of a hippocampal pyramidal neuron and the one in (B) to a striatal medium spiny neuron's.

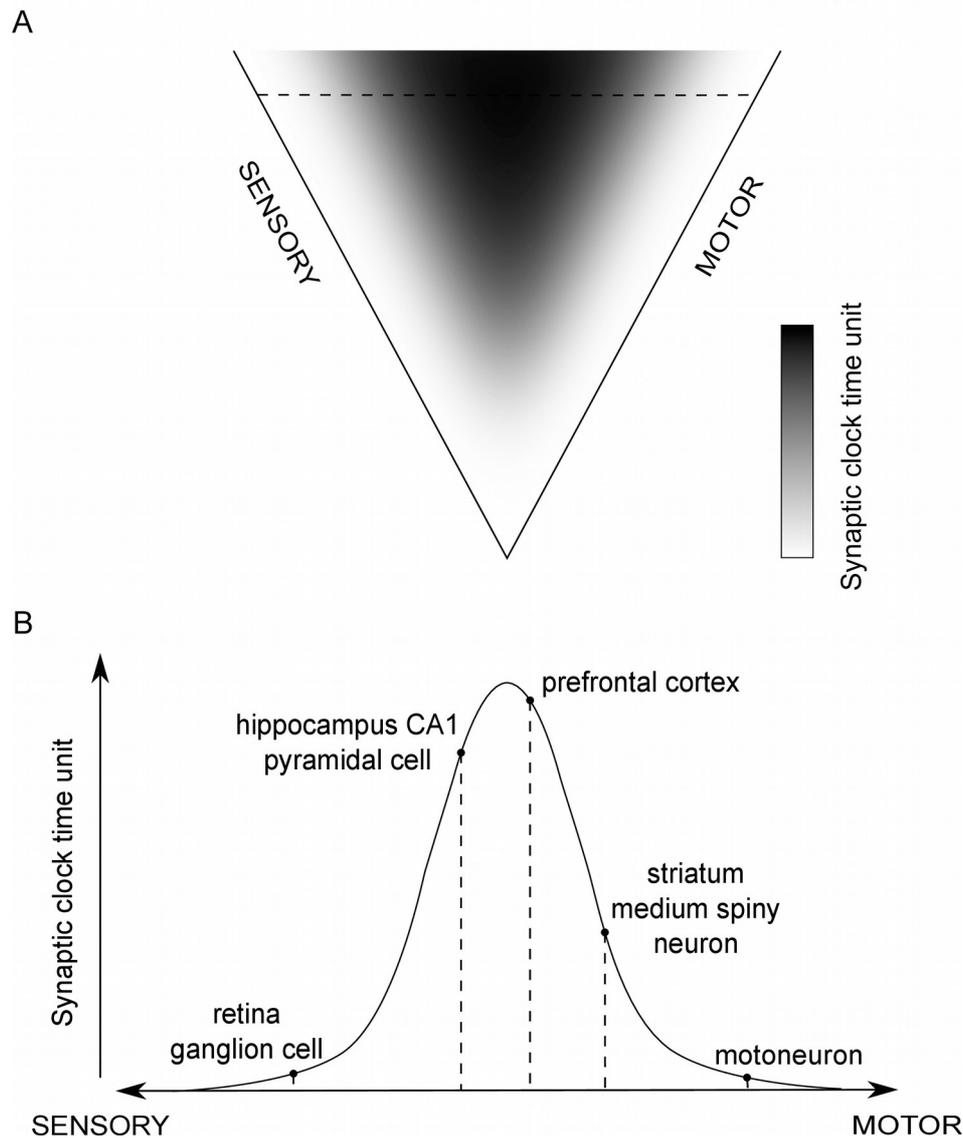

**Figure 4. A putative global distribution of synaptic clocks and their action affecting an overall temporal perception.**

    **(A)** A distribution of time units of different synaptic clocks, depending on their position on a sensory-motor continuum. Color-coded is the level of abstraction of information processed in a synapse, corresponding to the value of its time unit. **(B)** An example set of activated synapses with different time units, indicated in (A) as the horizontal dashed line. The pathway is going through the hippocampus, prefrontal cortex (PFC) and striatum. The resulting dynamics of synaptic clocks and the overall, averaged temporal perception might refer to a situation in which an animal is performing a motor turn into a certain direction (action-selection - striatum; motor action - motoneuron) on a crossroad, while navigating (visual processing - retina; spatial navigation - hippocampus) to a target location, as stored in the long-term reference memory (PFC).